\documentclass[12pt,english,draftcls, one column]{IEEEtran}
\usepackage[T1]{fontenc}
\usepackage[latin9]{inputenc}
\usepackage{amsthm}
\usepackage{amsmath}
\usepackage{amssymb}
\usepackage{graphicx}

\makeatletter

\providecommand{\tabularnewline}{\\}

\theoremstyle{plain}

\theoremstyle{plain}

\ifCLASSINFOpdf
\else
\fi
\usepackage{babel}

\makeatother

\usepackage{babel}
\providecommand{\propositionname}{Proposition}
\providecommand{\theoremname}{Theorem}

\begin{document}

\title{Performance Evaluation of \\Multiterminal Backhaul Compression for\\Cloud
Radio Access Networks}

\author{Seok-Hwan Park, Osvaldo Simeone, Onur Sahin and Shlomo Shamai (Shitz)
\thanks{S.-H. Park and O. Simeone are with the Center for Wireless Communications
and Signal Processing Research (CWCSPR), ECE Department, New Jersey
Institute of Technology (NJIT), Newark, NJ 07102, USA (email: \{seok-hwan.park,
osvaldo.simeone\}@njit.edu).

O. Sahin is with InterDigital Inc., Melville, New York, 11747, USA
(email: Onur.Sahin@interdigital.com).

S. Shamai (Shitz) is with the Department of Electrical Engineering,
Technion, Haifa, 32000, Israel (email: sshlomo@ee.technion.ac.il).%
}}
\maketitle
\begin{abstract}
In cloud radio access networks (C-RANs), the baseband processing of
the available macro- or pico/femto-base stations (BSs) is migrated
to control units, each of which manages a subset of BS antennas. The
centralized information processing at the control units enables effective
interference management. The main roadblock to the implementation
of C-RANs hinges on the effective integration of the radio units,
i.e., the BSs, with the backhaul network. This work first reviews
in a unified way recent results on the application of advanced multiterminal,
as opposed to standard point-to-point, backhaul compression techniques.
The gains provided by multiterminal backhaul compression are then
confirmed via extensive simulations based on standard cellular models.
As an example, it is observed that multiterminal compression strategies
provide performance gains of more than 60\% for both the uplink and
the downlink in terms of the cell-edge throughput.\end{abstract}
\theoremstyle{theorem}
\newtheorem{theorem}{Theorem}
\theoremstyle{proposition}
\newtheorem{proposition}{Proposition}
\theoremstyle{lemma}
\newtheorem{lemma}{Lemma}
\theoremstyle{corollary}
\newtheorem{corollary}{Corollary}
\theoremstyle{definition}
\newtheorem{definition}{Definition}
\theoremstyle{remark}
\newtheorem{remark}{Remark}

\section{Introduction\label{sec:Introduction}}

A promising architecture for next-generation wireless cellular systems
prescribes the separation of localized and distributed radio units
from remote and centralized information processing, or control, nodes.
This architecture is often referred to as a \emph{cloud radio access}
\textit{network} (C-RAN) \cite{Alcatel}\cite{China}. The centralization
of information processing afforded by C-RANs potentially enables effective
interference management at the geographical scale covered by the distributed
radio units. The main roadblock to the realization of this potential
hinges on the effective integration of the wireless interface provided
by the radio units with the backhaul network \cite{Biermann-et-al}.
Current solutions, which are the object of various standardization
efforts \cite{CPRI}, prescribe the use of standard analog-to-digital
conversion (ADC) techniques in the uplink and standard digital-to-analog
conversion (DAC) techniques in the downlink. With these standard solutions,
backhaul capacity limitations are known to impose a formidable bottleneck
to the system performance (see, e.g., \cite{IDT}).

In order to alleviate the performance bottleneck identified above,
recent efforts by a number of wireless companies have targeted the
design of more advanced \emph{backhaul compression} schemes. These
are based on various ad hoc combinations of ADC and DAC techniques
and proprietary \emph{point-to-point} compression algorithms (see,
e.g., \cite{Alcatel}). However, as it is well known from network
information theory, point-to-point techniques generally fail to achieve
the optimal performance in even the simplest multiterminal settings
\cite{ElGamal-Kim}. Recent works have hence explored the performance
of \textit{multiterminal}, as opposed to standard point-to-point,
backhaul compression techniques for the uplink \cite{Sanderovich}-\cite{WeiYu}
and the downlink \cite{Park-et-al:TSP} of C-RAN systems. In this
paper, we first review these works in Sec. \ref{sec:Distributed-Compression}
for the uplink and in Sec. \ref{sec:Multivariate-Compression} for
the downlink in a unified fashion. We then provide extensive simulation
results based on standard cellular models \cite{3GPP} to lend evidence
to the gains provided by multiterminal backhaul compression as compared
to standard point-to-point techniques in Sec. \ref{sec:Numerical-Results}.

\textit{Notation}: For random variables $X$, $Y$ and $Z$, we adopt
standard information-theoretic definitions for the mutual information
$I(X;Y)$, conditional mutual information $I(X;Y|Z)$, differential
entropy $h(X)$ and conditional differential entropy $h(X|Y)$ \cite{ElGamal-Kim}.
Given a sequence $X_{1},\ldots,X_{m}$, we define a set $X_{\mathcal{S}}=\{X_{j}|j\in\mathcal{S}\}$
for a subset $\mathcal{S}\subseteq\{1,\ldots,m\}$. For random vectors
$\mathbf{x}$ and $\mathbf{y}$, we define the following correlation
matrices $\mathbf{\Sigma}_{\mathbf{x}}=\mathbb{E}[\mathbf{x}\mathbf{x}^{\dagger}]$,
$\mathbf{\Sigma}_{\mathbf{x},\mathbf{y}}=\mathbb{E}[\mathbf{x}\mathbf{y}^{\dagger}]$
and $\mathbf{\Sigma}_{\mathbf{x}|\mathbf{y}}=\mathbb{E}[\mathbf{x}\mathbf{x}^{\dagger}|\mathbf{y}]$.

\begin{figure}
\centering\includegraphics[width=15cm,height=13.5cm]{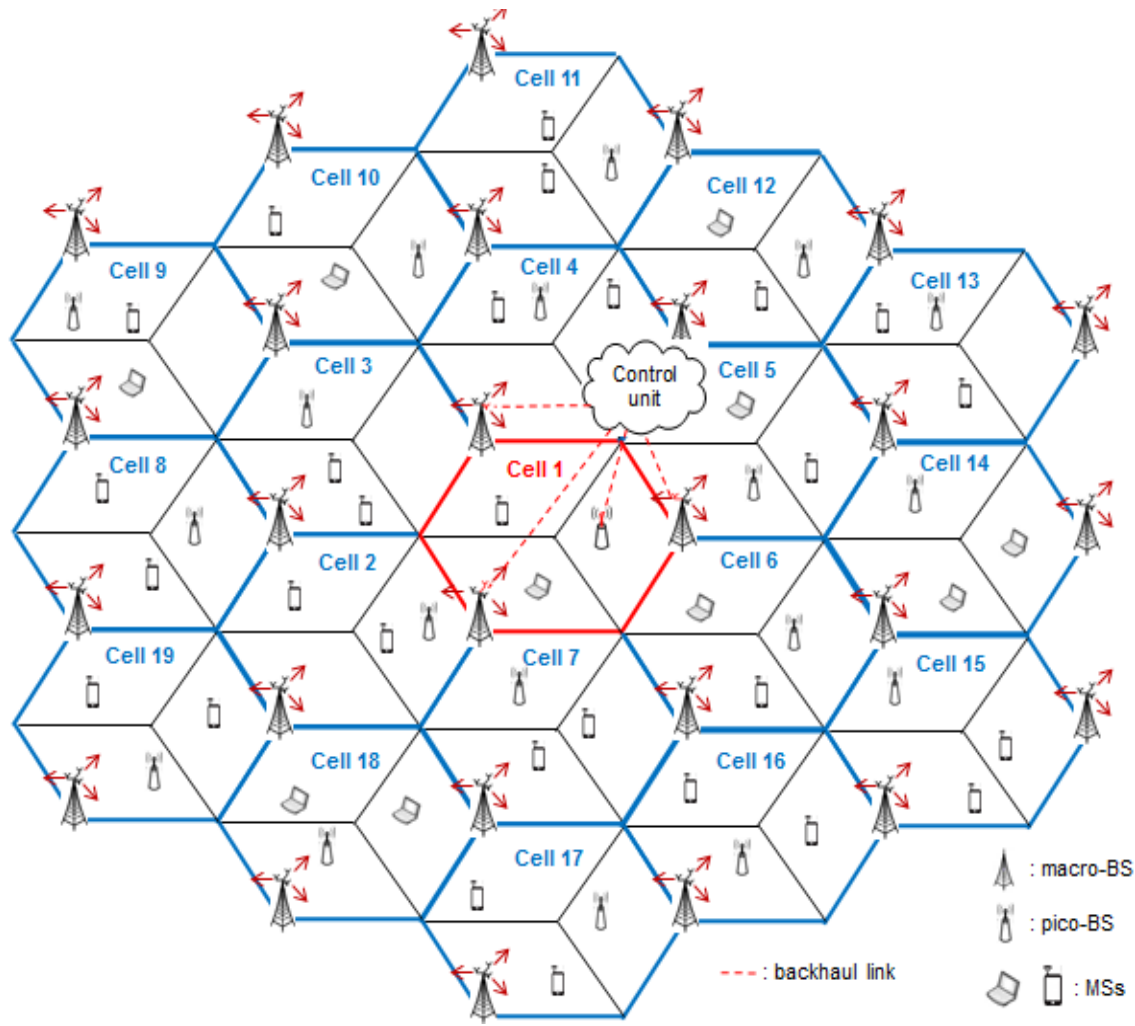}

\caption{\label{fig:layout-19}Two-dimensional hexagonal cellular layout with
19 macro hexagonal cells. Each macro BS has three sectorized antennas,
while pico-BSs and MSs use omni-directional antennas. We are interested
in the performance at macro cell 1 located at the center of the figure.}
\end{figure}

\section{System Model\label{sec:System-Model}}

We consider the two-dimensional hexagonal cellular layout with 19
macro cells shown in Fig. \ref{fig:layout-19}. We assume that each
macro-base station (BS) uses three sectorized antennas, and each pico-BS
and mobile station (MS) uses a single omni-directional antenna. In
each macro-cell, $K$ MSs and $N$ pico-BSs are uniformly distributed.
Fig. \ref{fig:layout-19} illustrates an example with $K=2$ MSs and
$N=1$ pico-BS.

In a C-RAN system, the baseband processing of the available macro-
or pico/femto-BSs is migrated to control units, each of which manages
a subset of BS antennas. For example, in Fig. \ref{fig:layout-19},
a control unit manages the three sectors of cell 1 and hence the corresponding
sectorial antennas of the three relevant macro-BSs and the available
pico-BS. We refer to a subset of BS antennas connected to the same
control unit, and to the corresponding covered area, as a \textit{cluster}.

Every $i$th BS is connected to the corresponding control unit via
a backhaul link with capacity $C_{i}$ bps/Hz \cite{Biermann-et-al}
where the normalization is done with respect to the bandwidth of the
wireless uplink/downlink channels. For instance, if BS $i$ communicates
with the corresponding control unit at a date rate of 100 Mbps and
the wireless uplink/downlink channels have a 10 MHz bandwidth, the
normalized backhaul capacity is given as $C_{i}=10$ bps/Hz.

According to the C-RAN principle, the data exchanged on the backhaul
links between BSs and control units consists of compressed baseband
signals \cite{Alcatel}-\cite{IDT}. Specifically, in the uplink, the
baseband signal received by each BS is compressed and forwarded to
the connected control unit, where decoding takes place. Instead, in
the downlink, the baseband signals are produced and compressed by
the control units, and then upconverted and transmitted by the BSs.

In the following, we detail the signal and channel model by focusing
on one specific cluster, e.g., cell 1 in Fig. \ref{fig:layout-19}.
For notational convenience, we index the BSs in the cluster as $1,2,\ldots,N_{B}$
and the MSs in the cluster as $1,2,\ldots,N_{M}$, and define the
sets $\mathcal{N_{B}}=\{1,\ldots,N_{B}\}$ and $\mathcal{N_{M}}=\{1,\ldots,N_{M}\}$.

\subsection{Uplink Channel\label{sub:Uplink-Channel}}

The signal $y_{i}^{\mathrm{ul}}$ received by BS $i$ in the cluster
under study in the uplink is given by
\begin{equation}
y_{i}^{\mathrm{ul}}=\mathbf{h}_{i}^{\mathrm{ul}\dagger}\mathbf{x^{\mathrm{ul}}}+z_{i}^{\mathrm{ul}},\label{eq:received-signal-BS}
\end{equation}
where $\mathbf{x^{\mathrm{ul}}=}[x_{1}^{\mathrm{ul}}\cdots x_{N_{M}}^{\mathrm{ul}}]^{T}$
is the $n_{M}\times1$ vector of symbols transmitted by all the $N_{M}$
MSs in the cluster, with $x_{k}^{\mathrm{ul}}$ being the symbol transmitted
by MS $k$; the noise $z_{i}^{\mathrm{ul}}\sim\mathcal{CN}(0,\sigma_{z_{i}^{\mathrm{ul}}}^{2})$
models thermal noise and the interference signals arising from the
other clusters; and the channel vector $\mathbf{h}_{i}^{\mathrm{ul}}\in\mathbb{C}^{N_{M}\times1}$
from all the $N_{M}$ MSs in the cluster toward BS $i$ is given by
$\mathbf{h}_{i}^{\mathrm{ul}}=[h_{i,1}^{\mathrm{ul}}\,\, h_{i,2}^{\mathrm{ul}}\,\,\cdots\,\, h_{i,N_{M}}^{\mathrm{ul}}]^{T}$
with $h_{i,k}^{\mathrm{ul}}$ denoting the uplink channel response
from the $k$th MS and to the $i$th BS. The signal $x_{k}^{\mathrm{ul}}$
is subject to the per-MS power constraint, which is stated as $\mathbb{E}[|x_{k}^{\mathrm{ul}}|^{2}]\leq P_{M,k}$
for $k\in\mathcal{N_{M}}$.

\subsection{Downlink Channel\label{sub:Downlink-Channel}}

In the downlink, each MS $k$ in the cluster under study receives
a signal given as
\begin{equation}
y_{k}^{\mathrm{dl}}=\mathbf{h}_{k}^{\mathrm{dl}\dagger}\mathbf{x}^{\mathrm{dl}}+z_{k}^{\mathrm{dl}},\label{eq:received-signal-MS}
\end{equation}
where we have defined the aggregate transmit signal vector by all
the $N_{B}$ BSs in the cluster as $\mathbf{x}^{\mathrm{dl}}=[x_{1}^{\mathrm{dl}},\ldots,x_{N_{B}}^{\mathrm{dl}}]^{T}$
with $x_{i}^{\mathrm{dl}}$ denoting the signal transmitted by the
$i$th BS; the additive noise $z_{k}^{\mathrm{dl}}\sim\mathcal{CN}(0,\sigma_{z_{k}^{\mathrm{dl}}}^{2})$
accounts for thermal noise and interference from the other clusters;
and the channel vector $\mathbf{h}_{k}^{\mathrm{dl}}\in\mathbb{C}^{N_{B}\times1}$
from all the BSs in the cluster toward MS $k$ is given as $\mathbf{h}_{k}^{\mathrm{dl}}=[h_{k,1}^{\mathrm{dl}}\,\, h_{k,2}^{\mathrm{dl}}\,\,\cdots\,\, h_{k,N_{B}}^{\mathrm{dl}}]^{T}$
with $h_{k,i}^{\mathrm{dl}}$ denoting the downlink channel gain from
BS $i$ to MS $k$. Finally, we have the per-BS power constraints
$\mathbb{E}[|x_{i}^{\mathrm{dl}}|^{2}]\leq P_{B,i}$, for $i\in\mathcal{N_{B}}$.

For both uplink and downlink, the channel vectors $\{\mathbf{h}_{i}^{\mathrm{ul}}\}_{i\in\mathcal{N_{B}}}$
and $\{\mathbf{h}_{k}^{\mathrm{dl}}\}_{k\in\mathcal{N_{M}}}$ remain
constant for the entire coding block duration and are known to the
corresponding control unit. As discussed in Sec. \ref{sec:Introduction},
the main goal of this paper is to provide a realistic evaluation of
the advantages of the multiterminal backhaul compression strategies
proposed in \cite{Park-et-al:TVT} for the uplink and in \cite{Park-et-al:TSP}
for the donwlink. In the next two sections, we review these strategies.

\section{Multiterminal Compression for the Uplink of C-RAN\label{sec:Distributed-Compression}}

\begin{figure}
\centering\includegraphics[width=13cm,height=10cm]{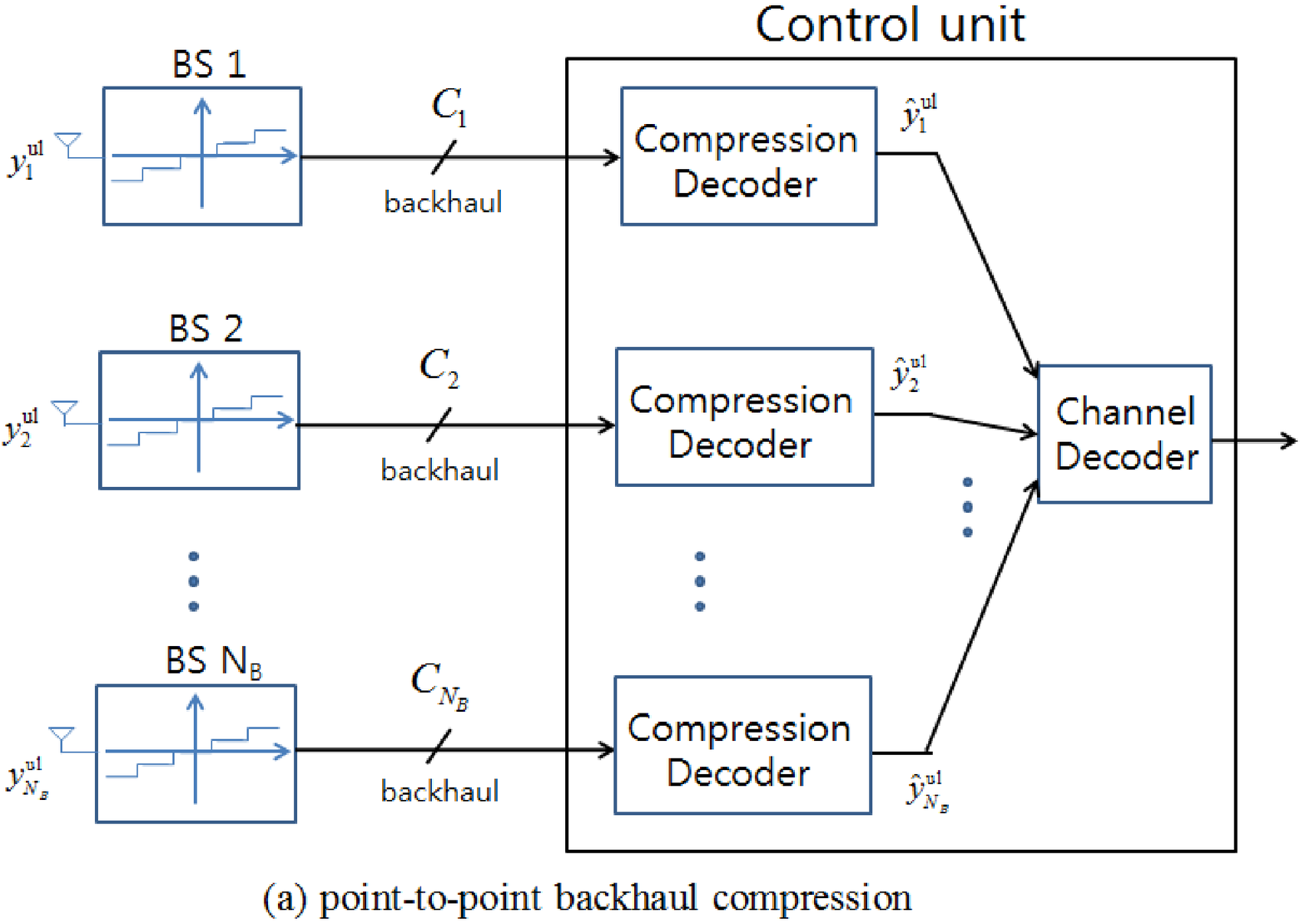}

\centering\includegraphics[width=14cm,height=9.7cm]{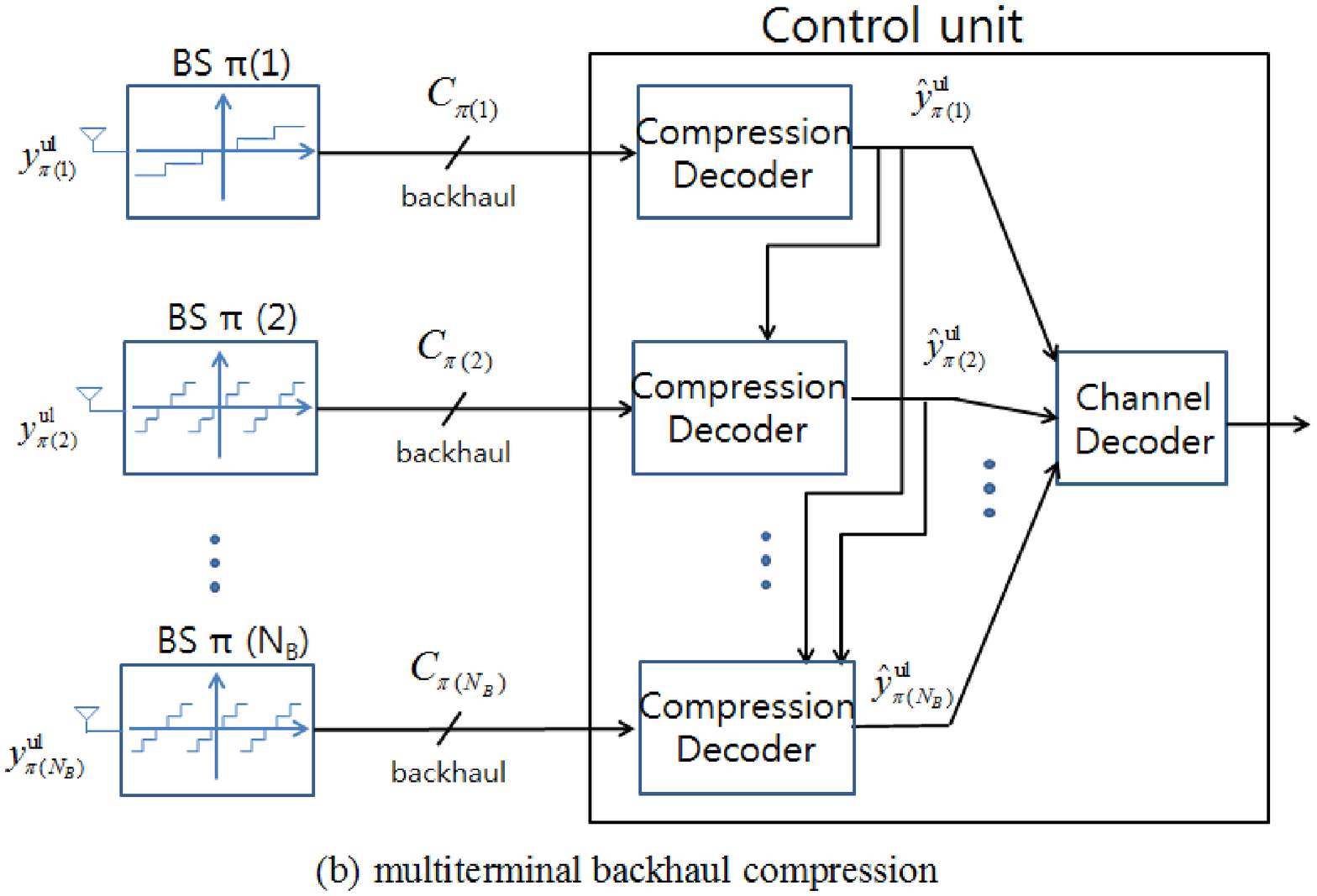}

\caption{\label{fig:UL-compression}Backhaul compression and decompression
for the uplink of C-RAN: (a) point-to-point compression; (b) multiterminal
compression.}
\end{figure}

In the \textit{uplink} of C-RAN, each MS $k$ within the cluster under
study encodes its message $M_{k}$ to produce a transmitted signal
$x_{k}^{\mathrm{ul}}$ for each channel use. This signal is taken
from a conventional Gaussian codebook and is hence distributed as
$x_{k}^{\mathrm{ul}}\sim\mathcal{CN}(0,P_{k})$ where $P_{k}$ satisfies
the per-MS power constraint $P_{k}\leq P_{M,k}$. Note that, since
the MSs cannot cooperate with each other, the transmitted signals
$x_{k}^{\mathrm{ul}}$ are independent across the MS index $k$.

Each $i$th BS communicates with the control unit by providing the
latter with a compressed version $\hat{y}_{i}^{\mathrm{ul}}$ of the
received signal $y_{i}^{\mathrm{ul}}$. The control unit first decompresses
the signals $\hat{y}_{i}^{\mathrm{ul}}$, $i\in\mathcal{N_{B}}$,
and then, based on all signals $\hat{y}_{\mathcal{N_{B}}}^{\mathrm{ul}}$,
decodes the MSs' messages%
\footnote{The advantage of joint decompression and decoding was studied in \cite{Park-et-al:SPL}.%
}. Using standard rate-distortion considerations, we express the compressed
signal $\hat{y}_{i}^{\mathrm{ul}}$ as%
\footnote{It is recalled that rate-distortion theory applies to vector quantizers
of large dimension although the mathematical characterizations of
the operation (such as (\ref{eq:Gaussian-compression-UL})) and of
the performance (such as (\ref{eq:UL-backhaul-indepQ}) below) are
given in terms of individual samples.%
}
\begin{equation}
\hat{y}_{i}^{\mathrm{ul}}=y_{i}^{\mathrm{ul}}+q_{i}^{\mathrm{ul}},\label{eq:Gaussian-compression-UL}
\end{equation}
where the quantization noise $q_{i}^{\mathrm{ul}}$ is independent
of the signal $y_{i}^{\mathrm{ul}}$ and distributed as $q_{i}^{\mathrm{ul}}\sim\mathcal{CN}(0,\omega_{i}^{\mathrm{ul}})$.

\textbf{Point-to-Point Backhaul Compression \cite{Fettweis}:} In
a conventional system, the control unit decompresses the descriptions
$\hat{y}_{\mathcal{N_{B}}}^{\mathrm{ul}}$ in parallel as shown in
Fig. \ref{fig:UL-compression}-(a). In this case, the signal $\hat{y}_{i}^{\mathrm{ul}}$
can be recovered at the control unit if the condition
\begin{align}
I(y_{i}^{\mathrm{ul}};\hat{y}_{i}^{\mathrm{ul}}) & =\log_{2}\left(\omega_{i}^{\mathrm{ul}}+\sigma_{y_{i}^{\mathrm{ul}}}^{2}\right)-\log_{2}\left(\omega_{i}^{\mathrm{ul}}\right)\leq C_{i}\label{eq:UL-backhaul-indepQ}
\end{align}
is satisfied where $\sigma_{y_{i}^{\mathrm{ul}}}^{2}=\mathbf{h}_{i}^{\mathrm{ul}\dagger}\mathbf{\Sigma}_{\mathbf{x}^{\mathrm{ul}}}\mathbf{h}_{i}^{\mathrm{ul}}+\sigma_{z_{i}^{\mathrm{ul}}}^{2}$
with $\mathbf{\Sigma}_{\mathbf{x}^{\mathrm{ul}}}=\mathrm{diag}(\{P_{k}\}_{k\in\mathcal{N_{M}}})$
(see, e.g., \cite[Ch. 3]{ElGamal-Kim}).

\textbf{Multiterminal Backhaul Compression \cite{Sanderovich}-\cite{WeiYu}:}
Standard point-to-point compression does not leverage the statistical
correlation among the signals $y_{i}^{\mathrm{ul}}$ received at different
BSs. Based on this observation, distributed compression was proposed
in \cite{Sanderovich} to utilize such correlation. Following \cite{Park-et-al:TVT}\cite{WeiYu},
this can be done as follows. For a given ordering $\pi$ of the BS
indices, the control unit decompresses in the order $\hat{y}_{\pi(1)}^{\mathrm{ul}},\hat{y}_{\pi(2)}^{\mathrm{ul}},\ldots,\hat{y}_{\pi(N_{B})}^{\mathrm{ul}}$
as shown in Fig. \ref{fig:UL-compression}-(b). Therefore, when decompressing
$\hat{y}_{\pi(i)}^{\mathrm{ul}}$, the control unit has already retrieved
the signals $\hat{y}_{\pi(1)}^{\mathrm{ul}},\ldots,\hat{y}_{\pi(i-1)}^{\mathrm{ul}}$.
These signals can be hence treated as side information available at
the decoder, namely the control unit, but not to the encoder, namely
BS $\pi(i)$. As a result, using the Wyner-Ziv theorem \cite[Ch. 3]{ElGamal-Kim},
the descriptions $\hat{y}_{\pi(i)}^{\mathrm{ul}}$ for $i\in\mathcal{N_{B}}$
can be recovered at the control unit if the conditions
\begin{align}
I(y_{\pi(i)}^{\mathrm{ul}};\hat{y}_{\pi(i)}^{\mathrm{ul}}|\hat{y}_{\{\pi(1),...,\pi(i-1)\}}^{\mathrm{ul}})= & g_{\pi,i}^{\mathrm{ul}}(\mathbf{p},\mbox{\boldmath\ensuremath{{\omega}}})\label{eq:backhaul-constraint-UL}\\
\triangleq & \log_{2}\left(\omega_{\pi(i)}^{\mathrm{ul}}+\sigma_{y_{\pi(i)}^{\mathrm{ul}}|\hat{y}_{\{\pi(1),...,\pi(i-1)\}}^{\mathrm{ul}}}^{2}\right)-\log_{2}\left(\omega_{\pi(i)}^{\mathrm{ul}}\right)\leq C_{\pi(i)}\nonumber
\end{align}
are satisfied, where we have defined vectors $\mathbf{p}=[P_{1},\ldots,P_{N_{M}}]$
and $\mbox{\boldmath\ensuremath{{\omega}}}=[\omega_{1},\ldots,\omega_{N_{B}}]$,
and the conditional variance $\sigma_{y_{\pi(i)}^{\mathrm{ul}}|\hat{y}_{\{\pi(1),...,\pi(i-1)\}}^{\mathrm{ul}}}^{2}$
is given by
\begin{equation}
\sigma_{y_{\pi(i)}^{\mathrm{ul}}|\hat{y}_{\{\pi(1),...,\pi(i-1)\}}^{\mathrm{ul}}}^{2}=\mathbf{h}_{\pi(i)}^{\mathrm{ul}\dagger}\mathbf{\Sigma}_{\mathbf{x}^{\mathrm{ul}}|\hat{y}_{\{\pi(1),...,\pi(i-1)\}}^{\mathrm{ul}}}\mathbf{h}_{\pi(i)}^{\mathrm{ul}}+\sigma_{z_{\pi(i)}^{\mathrm{ul}}}^{2},\label{eq:UL-conditional-variance}
\end{equation}
with $\mathbf{\Sigma}_{\mathbf{x}^{\mathrm{ul}}|\hat{y}_{\{\pi(1),...,\pi(i-1)\}}^{\mathrm{ul}}}=\mathbf{\Sigma}_{\mathbf{x}^{\mathrm{ul}}}-\mathbf{\Sigma}_{\mathbf{x}^{\mathrm{ul}},\hat{y}_{\{\pi(1),...,\pi(i-1)\}}^{\mathrm{ul}}}\mathbf{\Sigma}_{\hat{y}_{\{\pi(1),...,\pi(i-1)\}}^{\mathrm{ul}}}^{-1}\mathbf{\Sigma}_{\mathbf{x}^{\mathrm{ul}},\hat{y}_{\{\pi(1),...,\pi(i-1)\}}^{\mathrm{ul}}}^{\dagger}$.
The matrices $\mathbf{\Sigma}_{\mathbf{x}^{\mathrm{ul}},\hat{y}_{\{\pi(1),...,\pi(i-1)\}}^{\mathrm{ul}}}$
and $\mathbf{\Sigma}_{\hat{y}_{\{\pi(1),...,\pi(i-1)\}}^{\mathrm{ul}}}$
are given by
\begin{align}
 & \mathbf{\Sigma}_{\mathbf{x}^{\mathrm{ul}},\hat{y}_{\{\pi(1),...,\pi(i-1)\}}^{\mathrm{ul}}}=\mathbf{\Sigma}_{\mathbf{x}^{\mathrm{ul}}}\mathbf{H}_{\pi,i-1}^{\mathrm{ul}\dagger},\\
\mathrm{and}\,\, & \mathbf{\Sigma}_{\hat{y}_{\{\pi(1),...,\pi(i-1)\}}^{\mathrm{ul}}}=\mathbf{H}_{\pi,i-1}^{\mathrm{ul}}\mathbf{\Sigma}_{\mathbf{x}^{\mathrm{ul}}}\mathbf{H}_{\pi,i-1}^{\mathrm{ul}\dagger}+\mathrm{diag}\left(\{\sigma_{z_{\pi(j)}^{\mathrm{ul}}}^{2}+\omega_{\pi(j)}^{2}\}_{j=1}^{i-1}\right),
\end{align}
where we have defined the matrix $\mathbf{H}_{\pi,i-1}^{\mathrm{ul}}=[\mathbf{h}_{\pi(1)}^{\mathrm{ul}},\ldots,\mathbf{h}_{\pi(i-1)}^{\mathrm{ul}}]^{\dagger}$.

We assume that the control unit performs single-user decoding of the
messages $\{M_{k}\}_{k\in\mathcal{N_{M}}}$ sent by MSs based on all
the descriptions $\{\hat{y}_{i}^{\mathrm{ul}}\}_{i\in\mathcal{N_{B}}}$,
so that each message $M_{k}$ is decoded by treating the interference
signals $x_{j}^{\mathrm{ul}}$ for $j\neq k$ as noise (see \cite{Park-et-al:TVT}
for the analysis with joint decoding of all MSs and \cite{WeiYu}
for successive interference cancellation). Under this assumption,
the achievable rate $R_{k}$ for MS $k$ is given by
\begin{align}
R_{k}^{\mathrm{ul}}= & I(x_{k}^{\mathrm{ul}};\hat{y}_{\mathcal{N_{B}}}^{\mathrm{ul}})=f_{k}^{\mathrm{ul}}(\mathbf{p},\mbox{\boldmath\ensuremath{{\omega}}})\label{eq:rate-MS-k-UL}\\
\triangleq & \log_{2}\det\left(\mathrm{diag}\left(\{\omega_{i}^{\mathrm{ul}}\}_{i\in\mathcal{N_{B}}}\right)+\mathbf{\Sigma}_{\mathbf{y}_{\mathcal{N_{B}}}^{\mathrm{ul}}}\right)-\log_{2}\det\left(\mathrm{diag}\left(\{\omega_{i}^{\mathrm{ul}}\}_{i\in\mathcal{N_{B}}}\right)+\mathbf{\Sigma}_{\mathbf{y}_{\mathcal{N_{B}}}^{\mathrm{ul}}|x_{k}^{\mathrm{ul}}}\right),\nonumber
\end{align}
where the conditional covariance $\mathbf{\Sigma}_{\mathbf{y}_{\mathcal{N_{B}}}^{\mathrm{ul}}|x_{\mathcal{S}}^{\mathrm{ul}}}$
with $\mathcal{S}\subseteq\mathcal{N_{M}}$ is given as
\begin{equation}
\mathbf{\Sigma}_{\mathbf{y}_{\mathcal{N_{B}}}^{\mathrm{ul}}|x_{\mathcal{S}}^{\mathrm{ul}}}=\sum_{j\in\mathcal{N_{M}}\setminus\mathcal{S}}P_{j}\tilde{\mathbf{h}}_{j}^{\mathrm{ul}}\tilde{\mathbf{h}}_{j}^{\mathrm{ul}\dagger}+\mathrm{diag}\left(\{\sigma_{z_{i}^{\mathrm{ul}}}^{2}\}_{i\in\mathcal{N_{B}}}\right)
\end{equation}
with $\tilde{\mathbf{h}}_{k}^{\mathrm{ul}}=[h_{1,k}^{\mathrm{ul}},h_{2,k}^{\mathrm{ul}},\ldots,h_{N_{B},k}^{\mathrm{ul}}]^{T}$.

We are interested in evaluating the performance of the standard proportional-fair
scheduler. This scheduler, at each time slot, select the power allocation
$\mathbf{p}$ and the quantization noise powers $\mbox{\boldmath\ensuremath{{\omega}}}$
and the order $\pi$ so as to maximize the weighted sum-rate
\begin{equation}
u^{\mathrm{ul}}(\mathbf{p},\mbox{\boldmath\ensuremath{{\omega}}})=\sum_{k\in\mathcal{N_{M}}}f_{k}^{\mathrm{ul}}(\mathbf{p},\mbox{\boldmath\ensuremath{{\omega}}})/\bar{R}_{k}^{\alpha},\label{eq:alpha-fair-MS-k}
\end{equation}
with $\alpha\geq0$ being a fairness constant and $\bar{R}_{k}$ represents
the average data rate of MS $k$ until the previous time slot (see,
e.g., \cite{Viswanath-et-al}). After each time slot, the rate $\bar{R}_{k}$
is updated as $\bar{R}_{k}\leftarrow\beta\bar{R}_{k}+(1-\beta)R_{k}^{\mathrm{ul}}$
where $\beta\in[0,1]$ is a forgetting factor. We recall that increasing
the constant $\alpha$ encourages fairness among the MSs, while the
objective function reduces to the sum-rate when $\alpha=0$. This
problem is formulated as\begin{subequations}\label{eq:problem-UL}
\begin{align}
\underset{\pi,\mathbf{p}\in\mathbb{R}_{+}^{N_{M}},\mbox{\boldmath\ensuremath{{\omega}}}\in\mathbb{R}_{+}^{N_{B}}}{\mathrm{maximize}} & u^{\mathrm{ul}}(\mathbf{p},\mbox{\boldmath\ensuremath{{\omega}}})\label{eq:problem-UL-objective}\\
\mathrm{s.t.}\,\,\,\,\, & g_{\pi,i}^{\mathrm{ul}}(\mathbf{p},\mbox{\boldmath\ensuremath{{\omega}}})\leq C_{\pi(i)},\,\,\mathrm{for\,\, all}\,\, i\in\mathcal{N_{B}},\label{eq:problem-UL-backhaul}\\
 & P_{k}\leq P_{M,k},\,\,\mathrm{for\,\, all}\,\, k\in\mathcal{N_{M}}.\label{eq:problem-UL-power}
\end{align}
\end{subequations}To tackle the non-convex problem (\ref{eq:problem-UL}),
we propose a separate design of the power control variables $\mathbf{p}$
and the compression noise powers $\mbox{\boldmath\ensuremath{{\omega}}}$
for a fixed permutation $\pi$. Specifically, at Step 1, the power
coefficients $\mathbf{p}$ are optimized assuming ideal backhaul links
(i.e., $\mathbf{\omega}_{i}^{\mathrm{ul}}=0$ for $i\in\mathcal{N_{B}}$).
This problem is stated as
\begin{align}
\underset{\mathbf{p}\in\mathbb{R}_{+}^{N_{M}}}{\mathrm{maximize}} & \,\, u^{\mathrm{ul}}(\mathbf{p},\mathbf{0})\label{eq:problem-UL-reduced}\\
\mathrm{s.t.}\,\,\,\,\, & P_{k}\leq P_{M,k},\,\,\mathrm{for\,\, all}\,\, k\in\mathcal{N_{M}},\nonumber
\end{align}
or, equivalently, in the epigraph form\begin{subequations}\label{eq:epigraph-UL}

\begin{align}
\underset{R_{k},\mathbf{p}\in\mathbb{R}_{+}^{N_{M}}}{\mathrm{maximize}} & \,\,\sum_{k\in\mathcal{N_{M}}}R_{k}/\bar{R}_{k}^{\alpha}\label{eq:epigraph-UL-objective}\\
\mathrm{s.t.}\,\,\,\,\, & R_{k}\leq f_{k}^{\mathrm{ul}}(\mathbf{p},\mathbf{0}),\,\,\mathrm{for\,\, all}\,\, k\in\mathcal{N_{M}},\label{eq:epigraph-UL-rate}\\
 & P_{k}\leq P_{M,k},\,\,\mathrm{for\,\, all}\,\, k\in\mathcal{N_{M}}.\label{eq:epigraph-UL-power}
\end{align}
\end{subequations}Albeit still non-convex, it is seen that the problem
(\ref{eq:epigraph-UL}) belongs to the class of different-of-convex
(DC) problems (see, e.g., \cite{Beck}). Thus, we can leverage the
iterative majorization minimization (MM) algorithm, which is known
to converge to a locally optimal point of (\ref{eq:epigraph-UL})
(see, e.g., \cite[Sec. 1.3.3]{Beck}). The MM algorithm solves a sequence
of convex problems obtained by linearizing the non-convex constraints
(\ref{eq:epigraph-UL-rate}). With the so-obtained power variables
$\mathbf{p}$, at Step 2, we optimize the quantization noise powers
$\mbox{\boldmath\ensuremath{{\omega}}}$. It can be seen that the
optimal quantization power $\omega_{\pi(i)}^{\mathrm{ul}}$, for fixed
powers $\mathbf{p}$, is simply given by imposing equality in the
backhaul constraint (\ref{eq:problem-UL-backhaul}), leading to
\begin{equation}
\omega_{\pi(i)}^{\mathrm{ul}}=\sigma_{y_{\pi(i)}^{\mathrm{ul}}|\hat{y}_{\{\pi(1),...,\pi(i-1)\}}^{\mathrm{ul}}}^{2}/(2^{C_{\pi(i)}}-1)\label{eq:solution-UL-single-antenna}
\end{equation}
for $i\in\mathcal{N_{B}}$ with $\sigma_{y_{\pi(i)}^{\mathrm{ul}}|\hat{y}_{\{\pi(1),...,\pi(i-1)\}}^{\mathrm{ul}}}^{2}$
given in (\ref{eq:UL-conditional-variance}).

\section{Multiterminal Compression for the Downlink of C-RAN\label{sec:Multivariate-Compression}}

\begin{figure}
\centering\includegraphics[width=17cm,height=8cm]{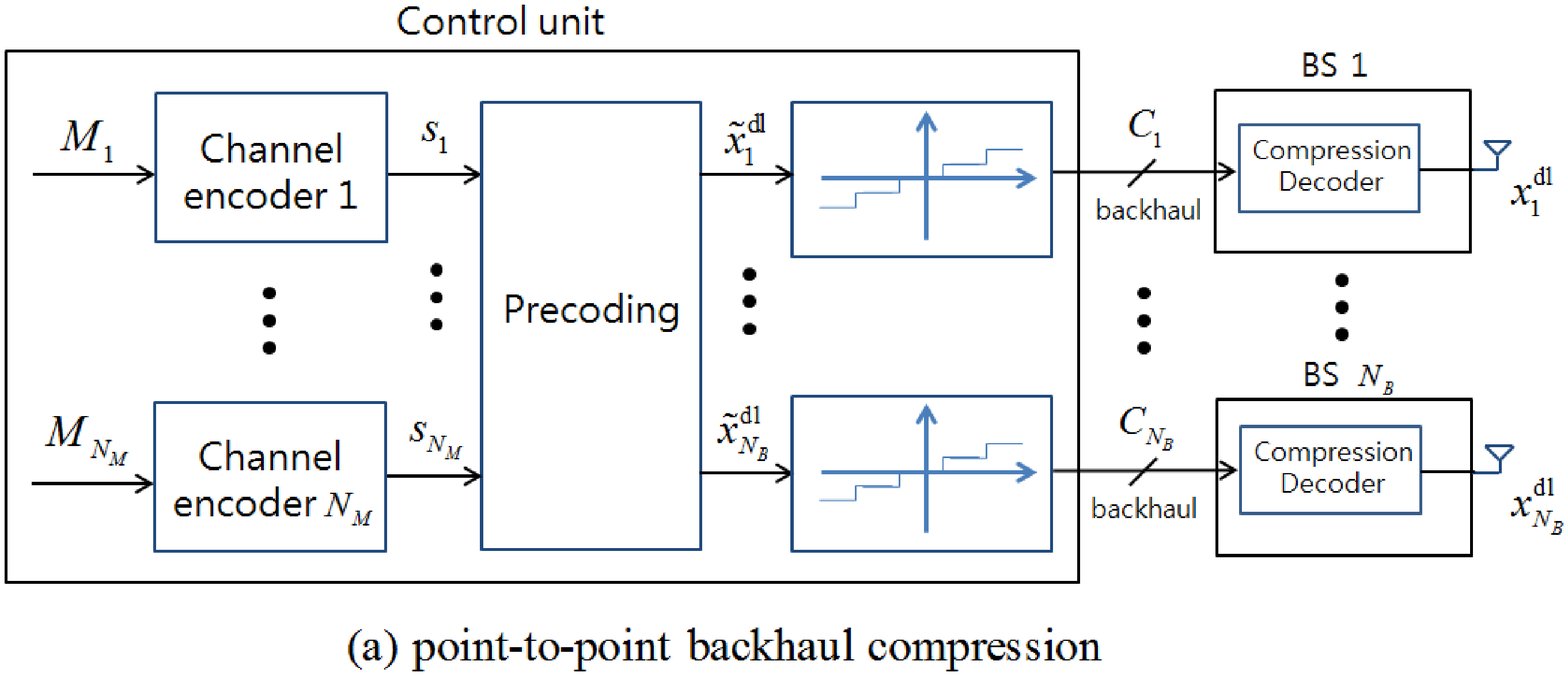}\\

\centering\includegraphics[width=17cm,height=7.3cm]{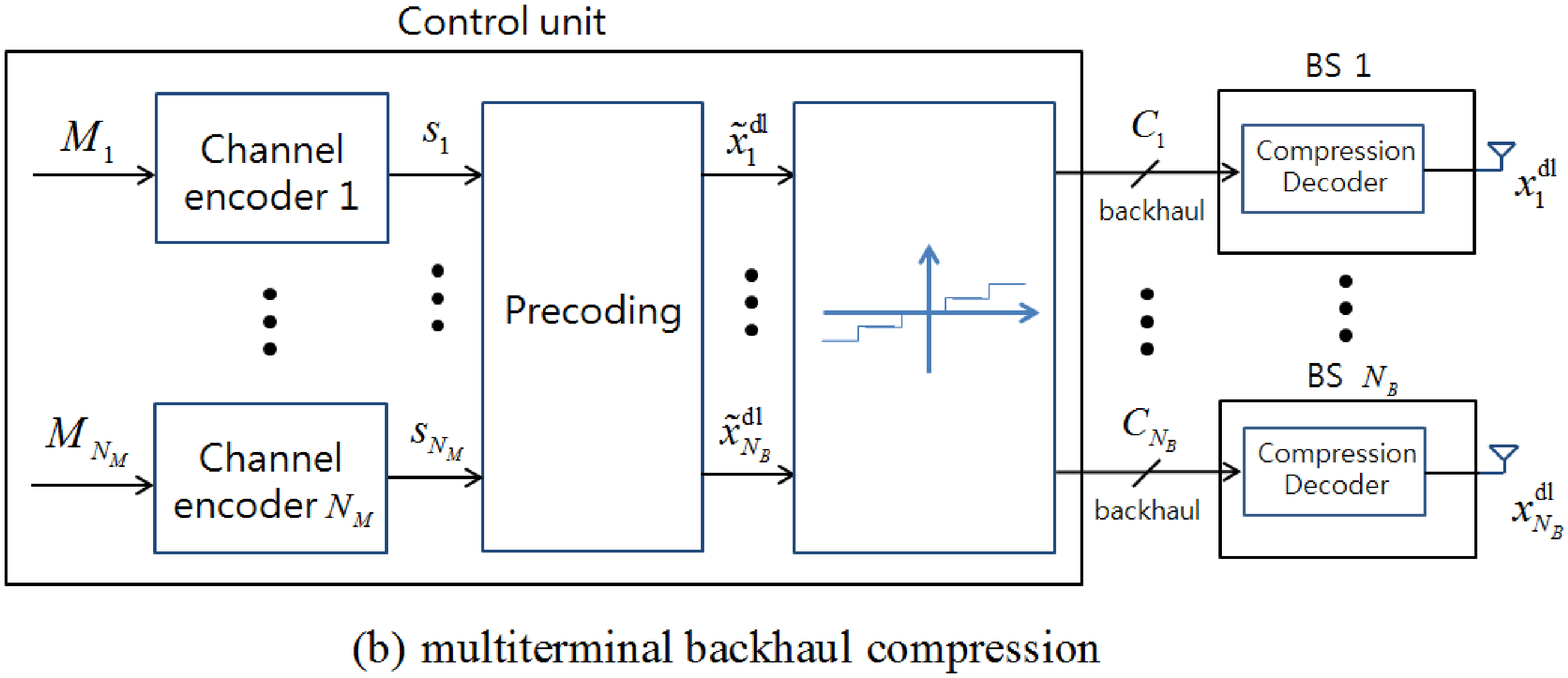}

\caption{\label{fig:DL-compression}Backhaul compression and decompression
for the downlink of C-RAN: (a) point-to-point compression; (b) multiterminal
compression.}
\end{figure}

In the \textit{downlink} of a C-RAN, the control unit first encodes
each message $M_{k}$ for MS $k\in\mathcal{N_{M}}$ via a separate
channel encoder, which produces a coded signal $s_{k}$ for each channel
use. Each coded symbol $s_{k}$ is taken from a conventional Gaussian
codebook and hence it is distributed as $s_{k}\sim\mathcal{CN}(0,1)$.
The signals $\mathbf{s}=[s_{1},\ldots,s_{N_{M}}]$ are further processed
by the control unit in two stages, namely \textit{precoding} and \textit{compression}.

\textbf{1. Precoding:} In order to allow for interference management
both across the MSs and among the data streams for the same MS, the
signals in vector $\mathbf{s}$ are linearly precoded via multiplication
of a complex matrix $\mathbf{A}\in\mathbb{C}^{N_{B}\times N_{M}}$.
The precoded data can be written as
\begin{equation}
\tilde{\mathbf{x}}^{\mathrm{dl}}=\mathbf{A}\mathbf{s},\label{eq:precoding-only}
\end{equation}
where the matrix $\mathbf{A}$ can be factorized as $\mathbf{A}=[\mathbf{a}_{1}\,\cdots\,\mathbf{a}_{N_{M}}]$
with $\mathbf{a}_{k}\in\mathbb{C}^{N_{B}\times1}$ denoting the precoding
vector corresponding to MS $k$. The precoded data $\tilde{\mathbf{x}}^{\mathrm{dl}}$
can be written as $\tilde{\mathbf{x}}^{\mathrm{dl}}=[\tilde{x}_{1}^{\mathrm{dl}},\ldots,\tilde{x}_{N_{B}}^{\mathrm{dl}}]^{T}$,
where the signal $\tilde{x}_{i}^{\mathrm{dl}}$ is the precoded signal
corresponding to the $i$th BS and is given as $\tilde{x}_{i}^{\mathrm{dl}}=\mathbf{e}_{i}^{\dagger}\mathbf{A}\mathbf{s}$
with the vector $\mathbf{e}_{i}\in\mathbb{C}^{N_{B}\times1}$ having
all zero elements except for the $i$th element that contains 1.

\textbf{2-(a). Point-to-Point Backhaul Compression \cite{Simeone}:}
Each precoded data stream $\tilde{x}_{i}^{\mathrm{dl}}$ for $i\in\mathcal{N_{B}}$
must be compressed in order to allow the control unit to deliver it
to the $i$th BS through the backhaul link of capacity $C_{i}$ bps/Hz.
Each $i$th BS then simply forwards the compressed signal $x_{i}^{\mathrm{dl}}$
obtained from the control unit. Using standard rate-distortion considerations,
we adopt a Gaussian test channel to model the effect of compression
on the backhaul link. In particular, we write the compressed signals
$x_{i}^{\mathrm{dl}}$ to be transmitted by BS $i$ as
\begin{align}
x_{i}^{\mathrm{dl}} & =\tilde{x}_{i}^{\mathrm{dl}}+q_{i}^{\mathrm{dl}},\label{eq:Gaussian-compression-DL-each-BS}
\end{align}
where the compression noise $q_{i}^{\mathrm{dl}}$ is modeled as a
complex Gaussian noise. With conventional backhaul compression, as
shown in Fig. \ref{fig:DL-compression}-(a), the signal $\tilde{x}_{i}^{\mathrm{dl}}$
corresponding to different BSs are compressed separately, which leads
to independent quantization noises $q_{i}^{\mathrm{dl}}$. Similar
to the uplink, the compressed signal (\ref{eq:Gaussian-compression-DL-each-BS})
can be transmitted to the $i$th BS if the condition
\begin{equation}
I\left(\tilde{\mathbf{x}}_{i};\mathbf{x}_{i}\right)=\log_{2}\left(\mathbf{e}_{i}^{\dagger}\mathbf{A}\mathbf{A}^{\dagger}\mathbf{e}_{i}+\omega_{i,i}^{\mathrm{dl}}\right)-\log_{2}\left(\omega_{i,i}^{\mathrm{dl}}\right)\leq C_{i}\label{eq:independent-comp-condition}
\end{equation}
is satisfied.

We now discuss the multiterminal backhaul compression strategies proposed
in \cite{Park-et-al:TSP}, and illustrated in Fig. \ref{fig:DL-compression}-(b).

\textbf{2-(b). Multiterminal Backhaul Compression \cite{Park-et-al:TSP}:}
The main idea of the multiterminal backhaul compression for the downlink
is to control the effect of the additive quantization noises at the
MSs by designing their correlation across the BSs within the cluster.
This is made possible by \textit{multivariate compression} \cite[Ch. 7]{ElGamal-Kim},
which requires joint compression of all signals as in Fig. \ref{fig:DL-compression}-(b).
A successive compression implementation, which is dual to the successive
decompression implementation of distributed source coding shown in
Fig. \ref{fig:UL-compression}-(b) for the uplink, is detailed in
\cite[Sec. IV-D]{Park-et-al:TSP}.

To elaborate, we write the vector $\mathbf{x}^{\mathrm{dl}}=[x_{1}^{\mathrm{dl}},\ldots,x_{N_{B}}^{\mathrm{dl}}]^{T}$
of compressed signals for all the BSs as
\begin{equation}
\mathbf{x}^{\mathrm{dl}}=\mathbf{A}\mathbf{s}+\mathbf{q}^{\mathrm{dl}}.\label{eq:whole-encoding-operation}
\end{equation}
In (\ref{eq:whole-encoding-operation}), the compression noise $\mathbf{q}^{\mathrm{dl}}=[q_{1}^{\mathrm{dl}},\ldots,q_{N_{B}}^{\mathrm{dl}}]^{T}$
is modeled as a complex Gaussian vector $\mathbf{q}^{\mathrm{dl}}\sim\mathcal{CN}(\mathbf{0},\mathbf{\Omega}^{\mathrm{dl}})$,
where the covariance matrix $\mathbf{\Omega}^{\mathrm{dl}}$ consists
of elements $\omega_{i,j}^{\mathrm{dl}}=\mathbb{E}[q_{i}^{\mathrm{dl}}q_{j}^{\mathrm{dl}\dagger}]$
defining the correlation between the quantization noises of BS $i$
and BS $j$.

Using the multivariate compression lemma in \cite[Ch. 9]{ElGamal-Kim},
reference \cite{Park-et-al:TSP} shows that the signals $x_{1}^{\mathrm{dl}},\ldots,x_{N_{B}}^{\mathrm{dl}}$
obtained via the test channel (\ref{eq:whole-encoding-operation})
can be reliably transferred to the BSs on the backhaul links if the
condition
\begin{align}
g_{\mathcal{S}}^{\mathrm{dl}}\left(\mathbf{A},\mathbf{\Omega}^{\mathrm{dl}}\right) & \triangleq\sum_{i\in\mathcal{S}}h\left(x_{i}^{\mathrm{dl}}\right)-h\left(x_{\mathcal{S}}^{\mathrm{dl}}|\tilde{\mathbf{x}}^{\mathrm{dl}}\right)\label{eq:multivariate-computed}\\
 & =\sum_{i\in\mathcal{S}}\log_{2}\left(\mathbf{e}_{i}^{\dagger}\mathbf{A}\mathbf{A}^{\dagger}\mathbf{e}_{i}+\omega_{i,i}^{\mathrm{dl}}\right)-\log_{2}\det\left(\mathbf{E}_{\mathcal{S}}^{\dagger}\mathbf{\Omega}^{\mathrm{dl}}\mathbf{E}_{\mathcal{S}}\right)\leq\sum_{i\in\mathcal{S}}C_{i}\nonumber
\end{align}
is satisfied for all subsets $\mathcal{S}\subseteq\mathcal{N_{B}}$,
where the matrix $\mathbf{E}_{\mathcal{S}}$ is obtained by stacking
the vectors $\mathbf{e}_{i}$ for $i\in\mathcal{S}$ horizontally.
We observe that the inequalities (\ref{eq:independent-comp-condition})
for standard point-to-point compression are obtained by substituting
$\omega_{i,j}^{\mathrm{dl}}=0$ into (\ref{eq:multivariate-computed}).

With the described precoding and compression operations and assuming
that the interference signals are treated as noise signals at MSs,
the achievable rate $R_{k}$ for MS $k$ is computed as
\begin{align}
R_{k} & =I\left(s_{k};y_{k}^{\mathrm{dl}}\right)=f_{k}^{\mathrm{dl}}\left(\mathbf{A},\mathbf{\Omega}^{\mathrm{dl}}\right)\label{eq:rate-MS-k-DL}\\
 & \triangleq\log_{2}\left(\sigma_{z_{k}^{\mathrm{dl}}}^{2}+\mathbf{h}_{k}^{\mathrm{dl}\dagger}\left(\mathbf{A}\mathbf{A}^{\dagger}+\mathbf{\Omega}^{\mathrm{dl}}\right)\mathbf{h}_{k}^{\mathrm{dl}}\right)-\log_{2}\left(\sigma_{z_{k}^{\mathrm{dl}}}^{2}+\mathbf{h}_{k}^{\mathrm{dl}\dagger}\left(\sum_{l\in\mathcal{N_{M}}\setminus\{k\}}\mathbf{a}_{l}\mathbf{a}_{l}^{\dagger}+\mathbf{\Omega}^{\mathrm{dl}}\right)\mathbf{h}_{k}^{\mathrm{dl}}\right).\nonumber
\end{align}

Similar to the uplink, our goal is to implement the proportional fairness
scheduler, which requires to optimize the weighted sum-rate over the
precoding matrix $\mathbf{A}$ and the quantization covariance matrix
$\mathbf{\mathbf{\Omega}}^{\mathrm{dl}}$, subject to the backhaul
constraints (\ref{eq:multivariate-computed}) and the per-BS power
constraints $P_{B,i}$. The weighted sum-rate $u^{\mathrm{dl}}(\mathbf{A},\mathbf{\Omega}^{\mathrm{dl}})$
is defined as in Sec. \ref{sec:Distributed-Compression}. This problem
is formulated as \begin{subequations}\label{eq:original-problem}
\begin{align}
\underset{\mathbf{A},\,\mathbf{\Omega}^{\mathrm{dl}}\succeq\mathbf{0}}{\mathrm{maximize}} & \,\, u^{\mathrm{dl}}(\mathbf{A},\mathbf{\Omega}^{\mathrm{dl}})\label{eq:original-problem-objective}\\
\mathrm{s.t.}\,\,\,\,\,\,\,\,\, & g_{\mathcal{S}}^{\mathrm{dl}}\left(\mathbf{A},\mathbf{\Omega}^{\mathrm{dl}}\right)\leq\sum_{i\in\mathcal{S}}C_{i},\,\,\mathrm{for\,\, all}\,\,\mathcal{S}\subseteq\mathcal{N_{B}},\label{eq:original-problem-backhaul}\\
 & \mathbf{e}_{i}^{\dagger}\mathbf{A}\mathbf{A}^{\dagger}\mathbf{e}_{i}+\omega_{i,i}^{\mathrm{dl}}\leq P_{B,i},\,\,\mathrm{for\,\, all}\,\, i\in\mathcal{N_{B}}.\label{eq:original-problem-power}
\end{align}
\end{subequations}A stationary point of problem (\ref{eq:original-problem})
can be found, as for (\ref{eq:problem-UL-reduced}), by applying the
MM algorithm on its epigraph form. The detailed algorithm can be derived
similar to \cite[Sec. V-A]{Park-et-al:TSP}.

\section{Performance Evaluation\label{sec:Numerical-Results}}

In this section, we discuss the performance advantages of multiterminal
backhaul compression for the uplink and downlink of C-RAN systems
on a standard cellular model based on \cite{3GPP}. We focus on the
performance evaluation in \textit{macro-cell 1} in Fig. \ref{fig:layout-19},
which is served by the three sectorized antennas from the corresponding
macro-BSs and by $N$ pico-BSs. A control unit is connected to all
BS antennas that serve cell 1 as in Fig. \ref{fig:layout-19}, which
is to be hence considered as a cluster. The backhaul links to each
macro-BS antenna and to each pico-BS have the capacities of $C_{\mathrm{macro}}$
and $C_{\mathrm{pico}}$ bps/Hz, respectively. All interference signals
from other macro-cells, denoted by cell 2, cell 3,$\ldots$, cell
19, are treated as independent noise signals. We used the system parameters
suggested in \cite{3GPP} and summarized in Table \ref{tab:parameters},
and adopted the LTE rate model proposed in \cite[Annex A]{3GPP-TR-136942}.
We assume that the fairness is measured during $T$ time slots in
which the locations of pico-BSs and MSs are fixed and small-scale
fading channels change independently from slot to slot.

\begin{table}
\centering%
\begin{tabular}{|c|c|}
\hline
Parameters & Assumptions\tabularnewline
\hline
\hline
System bandwidth & 10 MHz\tabularnewline
\hline
Path-loss (macro-BS) & $\mathrm{PL}\,(\mathrm{dB})=128.1+37.6\log_{10}R$, $R$: distance
in kilometers\tabularnewline
\hline
Path-loss (pico-BS) & $\mathrm{PL}\,(\mathrm{dB})=38+30\log_{10}R$, $R$: distance in meters\tabularnewline
\hline
Antenna pattern for sectorized macro-BS antennas & $A(\theta)=-\min\left[12(\theta/\theta_{3\mathrm{dB}})^{2},A_{m}\right]$,
$\theta_{3\mathrm{dB}}=65\textdegree$, $A_{m}=20\,\mathrm{dB}$\tabularnewline
\hline
Lognormal shadowing (macro-BS) & 10 dB standard deviation\tabularnewline
\hline
Lognormal shadowing (pico-BS) & 6 dB standard deviation\tabularnewline
\hline
Antenna gain after cable loss (macro-BS) & 15 dBi\tabularnewline
\hline
Antenna gain after cable loss (pico-BS and MS) & 0 dBi\tabularnewline
\hline
Noise figure & 5 dB (macro-BS), 6 dB (pico-BS), 9 dB (MS)\tabularnewline
\hline
Transmit power & 46 dBm (macro-BS), 24 dBm (pico-BS), 23 dBm (MS)\tabularnewline
\hline
\end{tabular}

\caption{{\footnotesize{\label{tab:parameters}Summary of the system parameters
used for simulation.}}}
\end{table}

\begin{figure}
\centering\includegraphics[width=10cm,height=9cm]{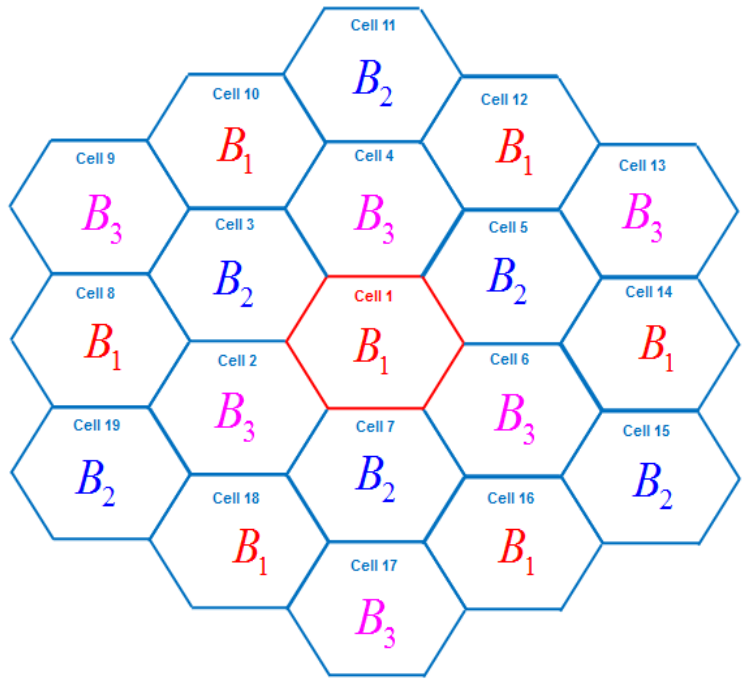}

\caption{\label{fig:frequency-reuse}Frequency reuse pattern proposed in \cite[Sec. VI]{Wang}
with reuse factor $F=1/3$ .}
\end{figure}

As shown in \cite{Wang}, with frequency reuse factor $F=1$, the
advantages of intra-cluster cooperation are masked by the effects
of the interference coming from the adjacent clusters. Thus, we consider
the frequency reuse pattern with $F=1/3$ proposed in \cite{Wang}
in which the available bandwidth is partitioned into three bands $B_{1}$,
$B_{2}$ and $B_{3}$, which are allocated so as to minimize the resulting
inter-cluster interference as illustrated in Fig. \ref{fig:frequency-reuse}.
As a result, cell 1 of interest suffers from the interference signals
only from cells (8,10,12,14,16,18).

\subsection{Uplink\label{sub:Uplink}}

\begin{figure}
\centering\includegraphics[width=12cm,height=9cm]{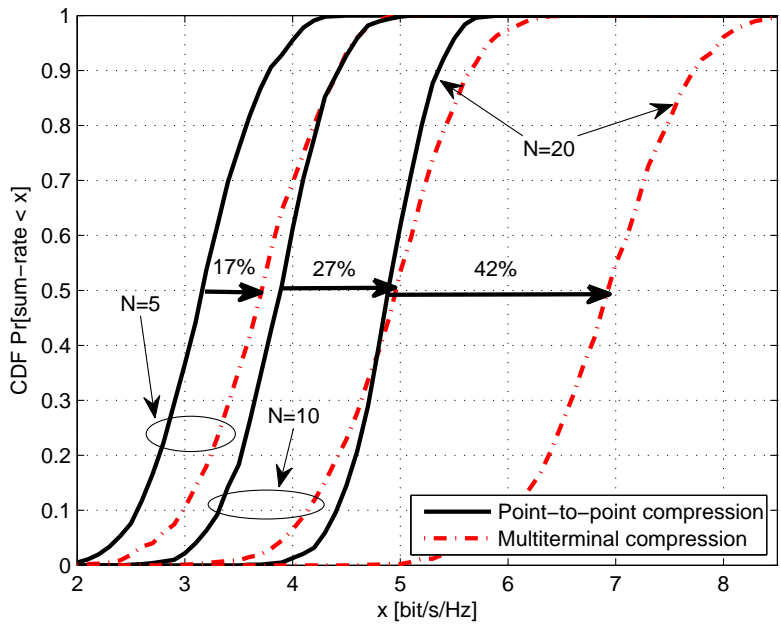}

\caption{\label{fig:graph-CDF-UL-FR}CDF of the sum-rate in the uplink C-RAN
with parameters as in \cite[Tables 5.3.3-1, 5.3.4-1]{3GPP}, $K=5$
MSs, $(C_{\mathrm{macro}},C_{\mathrm{pico}})=(3,1)$ bps/Hz and $\alpha=0$. }
\end{figure}

In this subsection, we examine the advantage of the multiterminal
compression scheme based on distributed source coding reviewed in
Sec. \ref{sec:Distributed-Compression} for the uplink of the C-RAN
described above. In Fig. \ref{fig:graph-CDF-UL-FR}, the CDF of the
sum-rate is plotted with $K=5$ MSs, $(C_{\mathrm{macro}},C_{\mathrm{pico}})=(3,1)$
bps/Hz and $\alpha=0$. For the order $\pi$ on the BS, we assume
that the control unit first retrieves the signals compressed at the
macro-BSs and then decompresses the signals received from the pico-BSs.
It is observed that, as compared to standard point-to-point compression,
multiterminal compression provides performance gains of 17\%, 27\%
and 42\% for $N=5$, $10$ and $20$ pico-BSs, respectively, in terms
of the 50\%-ile sum-rate. Thus, the performance gain of the multiterminal
compression is most pronounced when a large number of pico-BSs are
located in the same cluster. This suggests that a sophisticated design
of backhaul compression provides relevant gain if many radio units
are concentrated in given areas.

\begin{figure}
\centering\includegraphics[width=12cm,height=9cm]{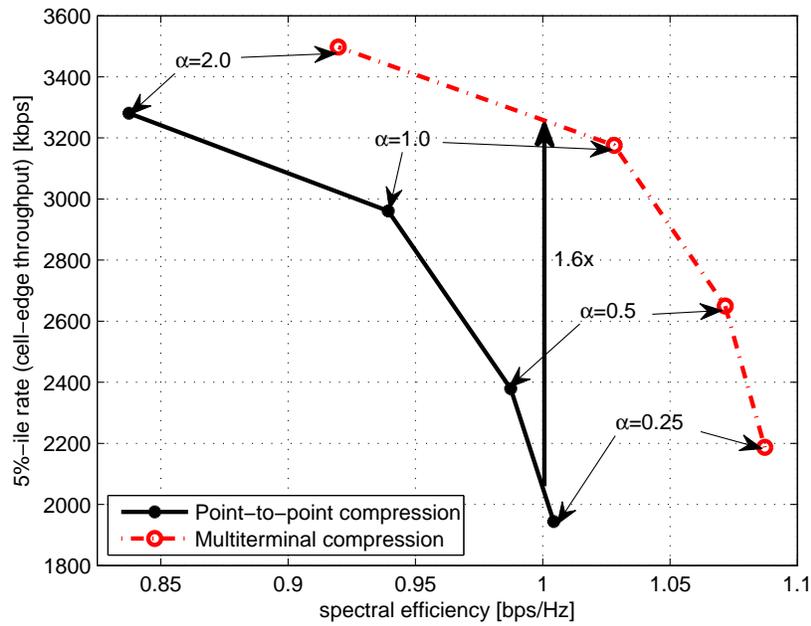}

\caption{\label{fig:graph-alpha-fair-UL}Cell-edge throughput, i.e., 5\%-ile
rate, versus the average spectral efficiency for various fairness
constants $\alpha$ in the uplink C-RAN with $N=3$ pico-BSs, $K=5$
MSs, $(C_{\mathrm{macro}},C_{\mathrm{pico}})=(9,3)$ bps/Hz, $T=10$
and $\beta=0.5$. }
\end{figure}

In Fig. \ref{fig:graph-alpha-fair-UL}, we plot the cell-edge throughput,
i.e., the 5\%-ile rate, versus the average spectral efficiency. The
curve is obtained by varying the fairness constant $\alpha$ in the
utility function (\ref{eq:alpha-fair-MS-k}) (see, e.g., \cite[Fig. 5]{Fettweis}).
We fix $N=3$ pico-BSs, $K=5$ MSs, $(C_{\mathrm{macro}},C_{\mathrm{pico}})=(9,3)$
bps/Hz, $T=10$ and $\beta=0.5$. As we increase the constant $\alpha$,
the 5\%-ile rate increases due to the enhanced fairness among the
MSs. We observe that spectral efficiencies larger than 1.01 bps/Hz
are not achievable with point-to-point compression, while they can
be obtained with multiterminal compression. Moreover, it is seen that
multiterminal compression provides 1.6x gain in terms of cell-edge
throughput for spectral efficiency of 2.9 bps/Hz.

\subsection{Downlink\label{sub:Downlink}}

\begin{figure}
\centering\includegraphics[width=12cm,height=9cm]{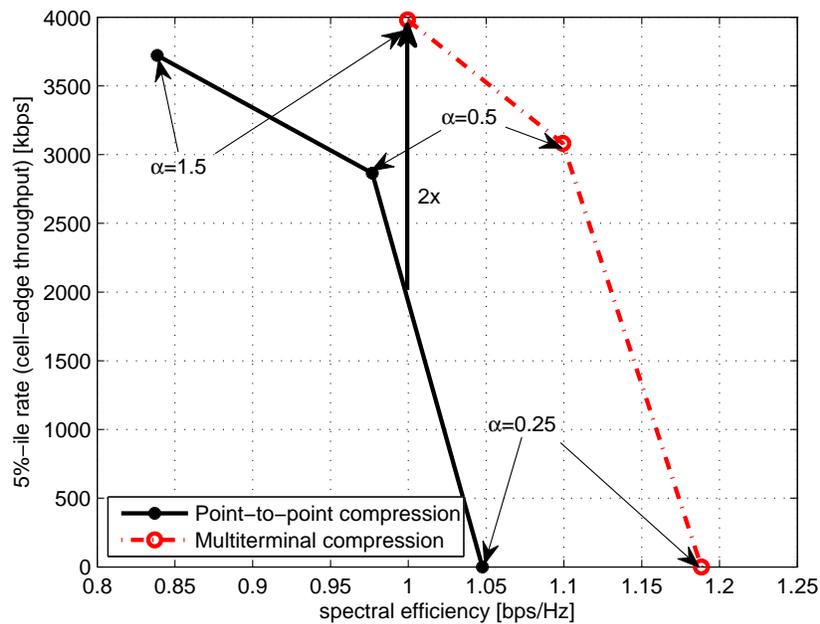}

\caption{\label{fig:graph-alpha-fair-DL}Cell-edge throughput, i.e., 5\%-ile
rate, versus the average spectral efficiency for various fairness
constants $\alpha$ in the downlink C-RAN with $N=1$ pico-BS, $K=4$
MSs, $(C_{\mathrm{macro}},C_{\mathrm{pico}})=(3,1)$ bps/Hz, $T=5$
and $\beta=0.5$. }
\end{figure}

In this subsection, we turn to the advantage of the multiterminal
compression technique as described in Sec. \ref{sec:Multivariate-Compression}
for the downlink. Fig. \ref{fig:graph-alpha-fair-DL} plots the cell-edge
throughput versus the average spectral efficiency for $N=3$ pico-BSs,
$K=5$ MSs, $(C_{\mathrm{macro}},C_{\mathrm{pico}})=(9,3)$ bps/Hz,
$T=10$ and $\beta=0.5$. As for the uplink, it is seen that spectral
efficiencies larger than 1.05 bps/Hz are not achievable with point-to-point
compression, while they can be obtained with multiterminal compression.
Specifically, multiterminal compression provides about 2x gain in
terms of cell-edge throughput for spectral efficiency of 1 bps/Hz.

\section{Conclusion\label{sec:Conclusion}}

In this work, we have studied the advantage of multiterminal backhaul
compression techniques over standard point-to-point compression for
the uplink and  downlink of cloud radio access networks. The extensive
simulations are based on standard cellular models and the results
focused on performance metrics such as sum-rate, proportional-fairness
utility and cell-edge throughput. As an example, we observed that
multiterminal compression techniques provide performance gains of
more than 60\% for both the uplink and the downlink in terms of the
cell-edge throughput.

\end{document}